# Adaptive Genetic Algorithm for Crystal Structure Prediction

**S Q Wu**[1,2], **M Ji**[2], **C Z Wang**[2], **M C Nguyen**[2], **X Zhao**[2], **K Umemoto**[2,3], **R M Wentzcovitch**[4] and **K M Ho**[2]
[1] Department of Physics, Xiamen University, Xiamen 361005, China
[2] Ames Laboratory - US DOE and Department of Physics and Astronomy, Iowa State University, Ames, Iowa 50011, USA
[3] Department of Earth Sciences, University of Minnesota, Minneapolis, MN 55455, USA
[4] Minnesota Supercomputing Institute and Department of Chemical Engineering and Materials Science, University of Minnesota, Minneapolis, MN 55455, USA

E-mail: wangcz@ameslab.gov (C Z Wang), kmh@ameslab.gov (K M Ho)

**Abstract.** We present a genetic algorithm (GA) for structural search that combines the speed of structure exploration by classical potentials with the accuracy of density functional theory (DFT) calculations in an adaptive and iterative way. This strategy increases the efficiency of the DFT-based GA by several orders of magnitude. This gain allows considerable increase in size and complexity of systems that can be studied by first principles. The method's performance is illustrated by successful structure identifications of complex binary and ternary inter-metallic compounds with 36 and 54 atoms per cell, respectively. The discovery of a multi-TPa Mg-silicate phase with unit cell containing up to 56 atoms is also reported. Such phase is likely to be an essential component of terrestrial exoplanetary mantles.

(Some figures may appear in colour only in the online journal)

---

Crystal structure prediction starting from the chemical composition alone has been one of the long-standing challenges in theoretical solid state physics, chemistry, and materials science [1,2]. Progress in this area has become a pressing issue in the age of computational materials discovery and design. In the past two decades several computational methods have been proposed to tackle this problem. These methods include simulated annealing [3-5], genetic algorithm (GA) [6-13], basin (or minima) hopping [14,15], particle swarm optimization [16,17], and *ab initio* random structure search [18]. While there has been steady progress in predicting crystal structures of elementary crystals, oxides, and binary alloys [8-13,16-18], exploration of complex binary, ternary, and quaternary systems has required more advanced algorithms for configuration space exploration and faster but reliable methods for energy evaluation. While first-principles density functional theory (DFT) calculations offer accurate total energies, its computational cost imposes the bottleneck to the structure identification of complex materials with unit cells containing ~$10^2$ atoms and/or with variable stoichiometries. By contrast, calculations based on classical potentials are fast and applicable to very large systems but are limited in accuracy. For various systems, reliable classical potentials are not even available. We present in this letter an *adaptive*-GA that combines the speed of classical potential searches and the accuracy of first-principles DFT calculations. It allows us to investigate crystal structures previously intractable by such methods with current computer capabilities.

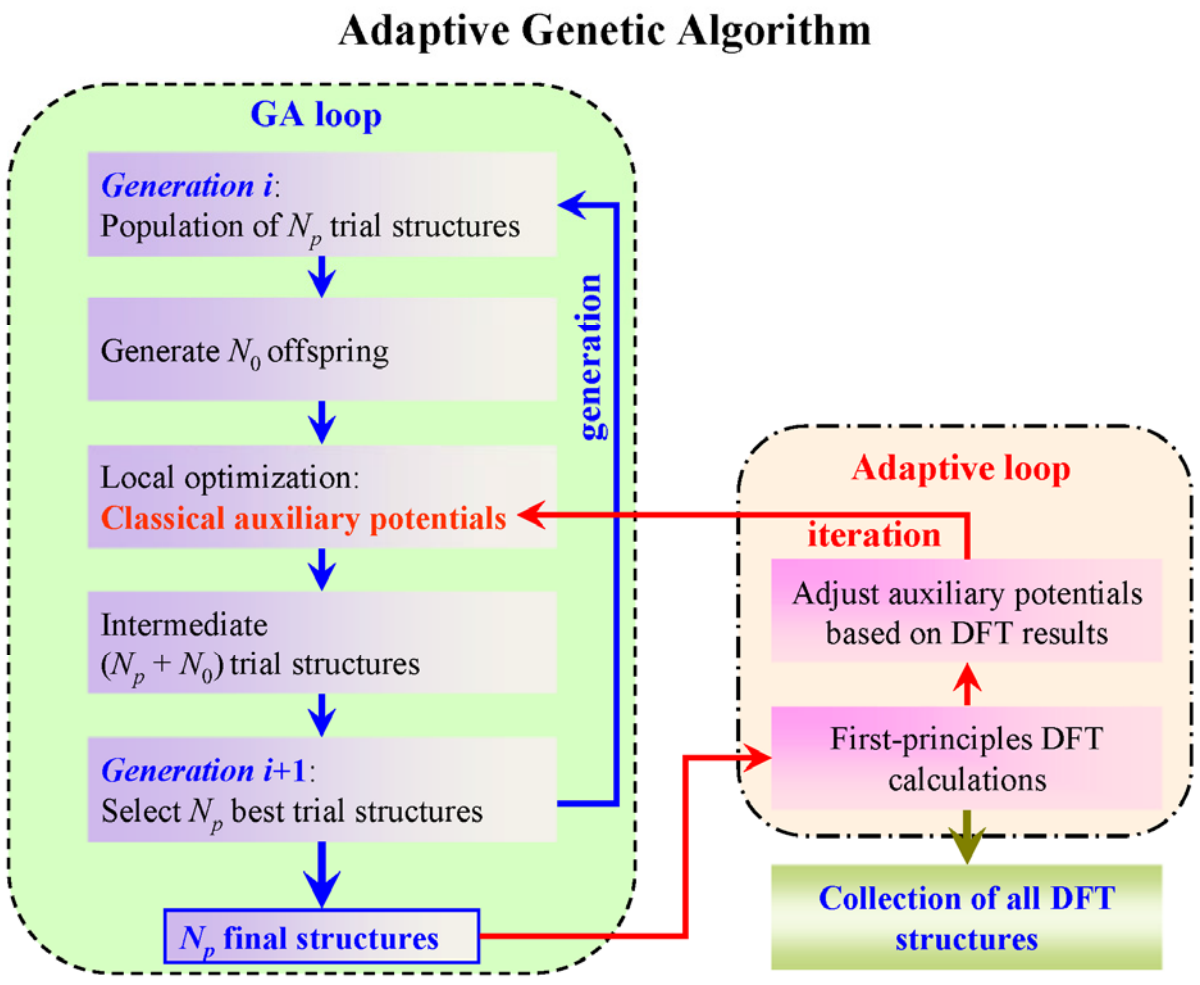

**Figure 1.** Flowchart of the adaptive genetic algorithm. The regular GA-loop is embedded in an adaptive loop. Optimization of offspring structures in the GA loop are performed using auxiliary classical potentials whose parameters are adjusted to reproduce DFT results obtained only in the adaptive loop.

The flowchart of the *adaptive*-GA scheme is illustrated in figure 1. The left-hand side of the flowchart is the traditional GA loop. The GA is an optimization strategy inspired by the Darwinian evolutionary process and has been widely adopted for atomistic structure optimization in the last 18 years [6-13]. During the GA optimization process, inheritance, mutation, selection, and crossover operations [6-13] are included to produce new structures and select most fit survivors from generation to generation. The most time-consuming step in the traditional GA-loop is the local optimization of new off-springs by DFT calculations. For complex structures, GA search usually iterates over 200 generations to converge. In the *adaptive*-GA scheme this most time-consuming step is performed using auxiliary classical potentials. In the *adaptive*-loop (see figure 1), single point DFT calculations are performed on a small set of candidate structures obtained in the GA-loop using the auxiliary classical potentials. Energies, forces, and stresses of these structures from first-principles DFT calculations are used to update the parameters of the auxiliary classical potentials by force-matching method with stochastic simulated annealing algorithm as implemented in the *potfit* code [19,20]. Another cycle of GA search is performed using the newly adjusted potentials, followed by the re-adjustment of the potential parameters, and the process is then repeated – an adaptive-GA (AGA) iteration. All first-principles DFT calculations were performed using the Quantum-ESPRESSO [21] or VASP [22,23], which has been interfaced with the adaptive-GA scheme in a fully parallel manner.

The numbers of parent, $N_p$, and off-spring structures, $N_o$, depend on the complexity of the system investigated. For those investigated here, $60 < N_o < 200$, and the total number of structures optimized in each GA-cycle varied between ~12,000 to ~40,000. The use of classical auxiliary potentials for such structure relaxations reduced the computational load by approximately five to six orders of magnitude. It usually takes 30-50 adaptive-GA-iterations to obtain the final structures and the net computational time of the entire adaptive-GA search can be reduced by more than three orders of magnitude. Since the classical potentials are adjusted according to DFT results, the adaptive-GA can explore configuration space more effectively. Structures collected over all adaptive-GA iterations and a set of low-energy metastable structures can be finally screened to locate the ground-state crystal structure. Therefore, the adaptive-GA can essentially search for structures almost with the efficiency of classical potentials but with DFT accuracy.

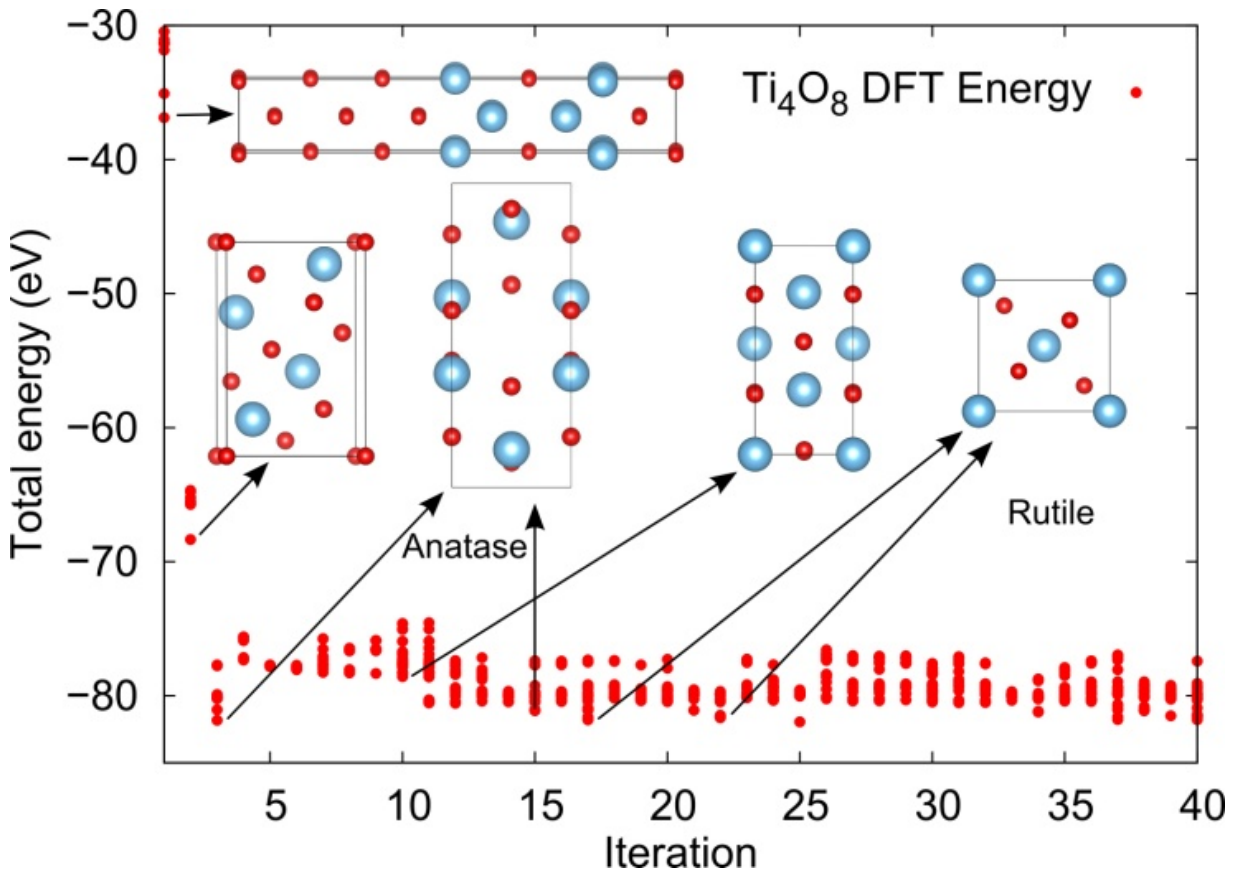

**Figure 2.** Structural and energetic evolution of $TiO_2$ versus iteration number of the AGA-loop. EAM-type potentials were used in inner GA-loop. Plots show only DFT energies obtained at the end of each AGA-iteration. The two low-energy crystal structures of $TiO_2$, *i.e.*, the rutile and the anatase structures could be obtained within 25 AGA-iterations.

We note that the commonly adopted approach of combining classical potentials with DFT calculations for structure optimization involves the use of a *single* set of classical potentials to screen *all* candidate structures followed by a refinement using DFT calculations. This requires accurate and transferable classical potentials able to capture the very-lowest (or few low) energy structures in a complex energy landscape. In contrast, from the energies of the final structures at each iteration as plotted in Fig 2 (and Fig 3 in the following), we can see that the adaptive-GA uses different adjusted potentials to sample structures located in different basins of the energy landscape. Each auxiliary classical potential may not just sample the structures in the same basin, it can sample the structures in a subset of the basins in the energy landscape and some of the basins may overlap with those from other potentials. Therefore adaptive GA is not designed to fit transferable potentials for general atomistic simulations. It is very difficult or even impossible to fit a classical potential able to accurately describe a system under various bonding environments, especially for binary and ternary systems. However, it is possible to adjust auxiliary potentials to describe structures located within different subset of basins in the energy landscape with DFT accuracy. Adapted auxiliary potentials adjusted throughout the adaptive-GA-iterations help the system to hop between basins and ensure efficient and accurate sampling of configuration space. An illustration of this point is the search for the crystal structure of $TiO_2$ with Embedded-atom method (EAM) type potentials. We do not expect EAM potentials to describe well the energies of various $TiO_2$ polymorphs. However, as seen in figure 2, the adaptive-GA search for structures with 4 formula units (12 atoms per unit cell) was able to find the two low-energy structures of $TiO_2$, *i.e.*, the rutile [24] and the anatase [25] structures – both with 6 atoms per primitive cell only – within 25 adaptive-GA-iterations. The theoretical structural parameters of the rutile and the anatase $TiO_2$ together with the experimental values are given in table 1.

**Table 1.** Structural parameters of Rutile & Anatase $TiO_2$

| $TiO_2$ | | Theo. | Expt.[24] |
|---|---|---|---|
| | | Rutile | |
| Space group | | *P*4$_2$/*mnm* | |
| ($a$, $c$) (Å) | | (4.6501, 2.9697) | (4.5929, 2.9591) |
| Ti | 2a | (0.0000, 0.0000, 0.0000) | (0, 0, 0) |
| O | 4f | (0.3049, 0.3049, 0.0000) | (0.3056, 0.3056, 0) |
| | | Anatase | |
| Space group | | *I*4$_1$/*amd* | |
| ($a$, $c$) (Å) | | (3.8074, 9.7050) | (3.785, 9.514) |
| Ti | 4a | (0.0000, 0.0000, 0.0000) | (0, 0, 0) |
| O | 8e | (0.0000, 0.0000, 0.2067) | (0, 0, 0.2064) |

To validate the adaptive-GA for complex crystal structure prediction we searched for structures of the $Hf_2Co_7$ binary alloy with 36 atoms per unit cell and of the $ZrCo_3B_2$ ternary alloy with 54 atoms per unit cell. Ground-state structures of these alloys have been well characterized experimentally [26, 27]. The adaptive-GA searches were performed using only chemical compositions as input information. Initial parent structures were generated by placing atoms randomly in the unit cell. EAM-type potentials were used in the GA-loop. As shown in figure 3, initially the unit cell shapes, atomic positions, and energies of $Hf_2Co_7$ and $ZrCo_3B_2$ structures are far from the ground-state. Adaptive-GA searches then quickly locate the correct ground-state structures for these alloys within 8 and 15 iterations, respectively. The predicted and experimental structural parameters for these two structures are presented in table 2 & 3. Note that for the $ZrCo_3B_2$, once the experimental structure (*R*-3) is relaxed by DFT method, a higher symmetry (*R*-3*m*) can be obtained. The *R*-3*m* structure is the ground state structure predicted by our adaptive-GA searches. These results demonstrate the power of the adaptive-GA as a computational tool for predicting complex crystal structures.

**Table 2.** Structural parameters of $Hf_2Co_7$

| $Hf_2Co_7$ | | Theo. | Expt.[26] |
|---|---|---|---|
| Space group | | *C*2/*m* | |
| (*a*, *b*, *c*) (Å) | | (4.6513, 8.2302, 12.0853) | (4.444, 8.191, 12.14) |
| *β* (º) | | 84.4110 | 90 |
| Hf(1) | 4i | (0.2187, 0.0000, 0.8830) | (0.27, 0, 0.884) |
| Hf(2) | 4i | (0.2857, 0.0000, 0.6172) | (0.212, 0, 0.613) |
| Co(1) | 8j | (-0.0025, 0.2499, 0.2499) | (0.053, 0.246, 0.251) |
| Co(2) | 8j | (0.2953, 0.1663, 0.0769) | (0.208, 0.162, 0.076) |
| Co(3) | 8i | (0.2006, 0.1664, 0.4232) | (0.297, 0.168, 0.421) |
| Co(4) | 4i | (0.2474, 0.0000, 0.2502) | (0.256, 0, 0.246) |

**Table 3.** Structural parameters of $ZrCo_3B_2$

| $ZrCo_3B_2$ | | Theo. | Expt.[27] |
|---|---|---|---|
| Space group | | *R*-3*m* | *R*-3 |
| (*a*, *c*) (Å) | | (8.4171, 9.0415) | (8.418, 9.132) |
| Zr(1) | 3a | (0.0000, 0.0000, 0.0000) | (0, 0, 0) |
| Zr(2) | 6c | (0.0000, 0.0000, 0.3304) | (0, 0, 0.328) |
| Co(1) | 9d | (0.5000, 0.0000, 0.5000) | (0.5, 0, 0.5) |
| Co(2) | 18h/18f | (0.1674, 0.8326, 0.1387) | (0.169, 0.834, 0.136) |
| B | 18f | (0.3756, 0.0000, 0.0000) | (0.072, 0.727, 0.33) |

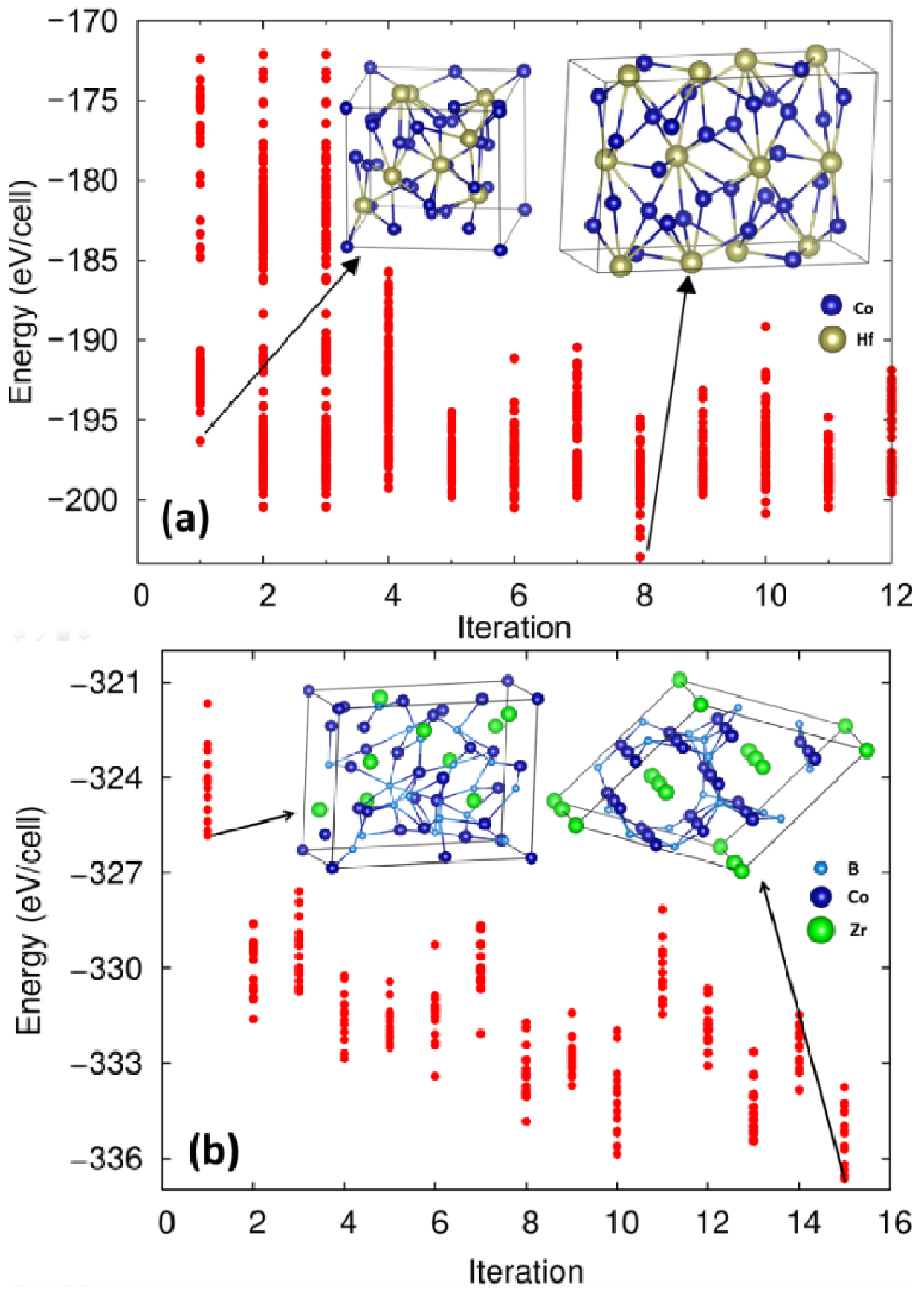

**Figure 3.** Structural and energetic evolution versus iteration number of the AGA-loop for (a) $Hf_2Co_7$ and (b) $ZrCo_3B_2$ alloys. EAM-type potentials were used in inner GA-loop. Plots show only DFT energies obtained at the end of each AGA-iteration.

Materials discovery is a very active area of research in mineral physics. This is motivated by the discovery of a myriad of candidates of "super-Earths" whose densities are comparable to Earth's and masses are 1~10 times Earth's mass by Kepler [28] and CoRot [29] missions. $MgSiO_3$ perovskite (Pv) is the most abundant mineral phase in Earth's lower mantle. In 2004, a post-perovskite phase (PPv) was stabilized by subjecting $MgSiO_3$-Pv to ~ 125 GPa and ~2,500 K [30]. The Pv→PPv transition [30-32] in $MgSiO_3$ seems to be associated with a major seismic discontinuity in the mantle at ~250 km above the core mantle boundary – the D" discontinuity. Although the PPv phase is the final form of $MgSiO_3$ in the Earth, further phase transitions are expected to occur in super-Earths. The dissociation of $MgSiO_3$-PPv into elementary oxides, $SiO_2$ and MgO, was then predicted to take place at ~1.1 TPa [33]. A more gradual dissociation process, a two-step ex-solution of MgO from $MgSiO_3$, has recently been identified [34]:

1) **$MgSiO_3$** (PPv) → **$MgSi_2O_5$** ($P2_1/c$-type) + **MgO** (CsCl-type)

2) **$MgSi_2O_5$** ($P2_1/c$-type) + **MgO** (CsCl-type) → **MgO** (CsCl-type) + **$SiO_2$** ($Fe_2P$-type)

This has raised new questions about the complexity of the dissociation process. Clarification of this process and of the nature of intermediate aggregates is essential for understanding the internal structure of terrestrial-type exoplanets. It is also relevant for advancing knowledge of the dense cores of the giant planets, which should contain these most abundant Earth forming elements in condensed form.

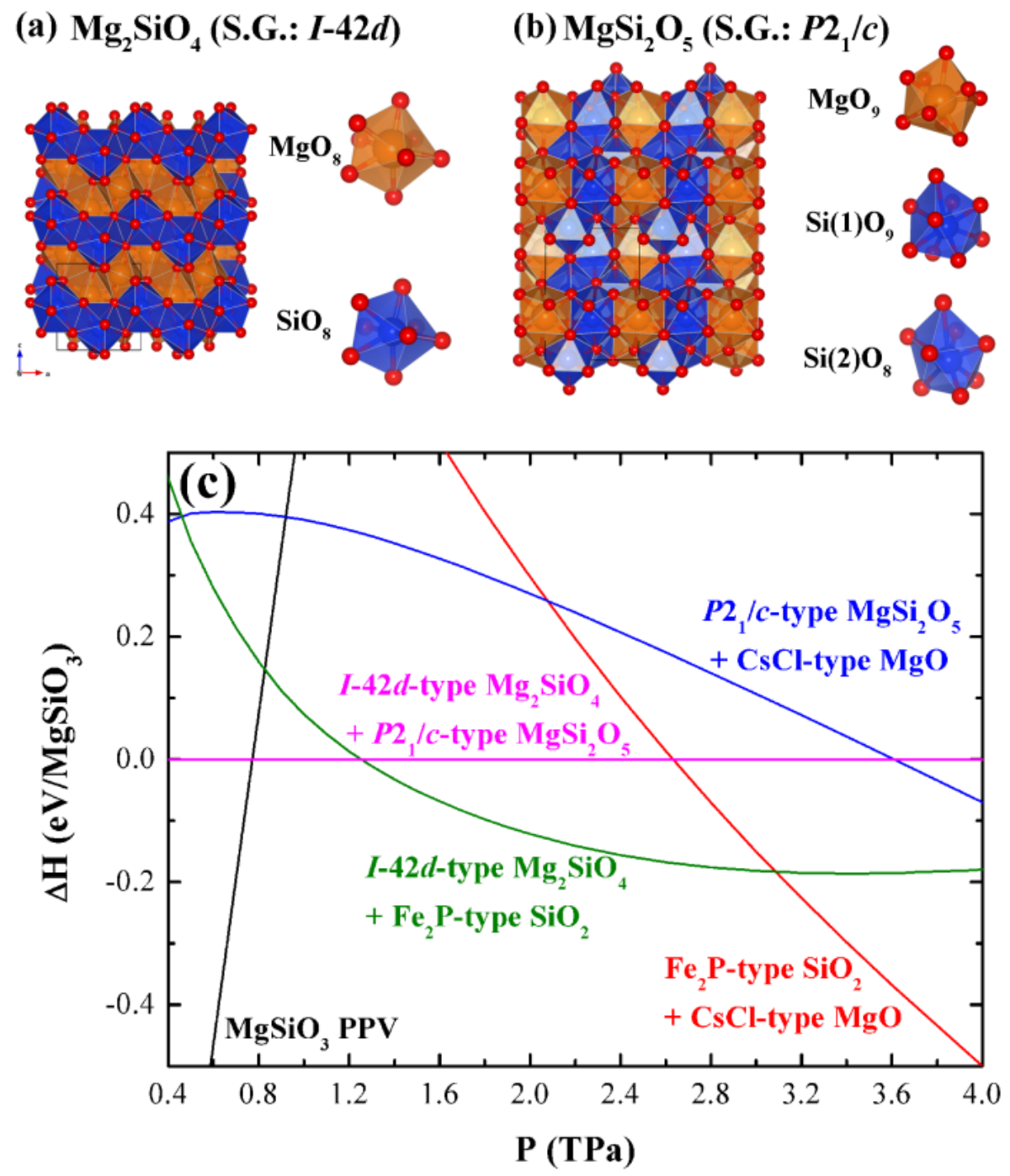

**Figure 4.** Crystal structures of (a) $I$-42$d$ $Mg_2SiO_4$ and (b) $P2_1/c$ $MgSi_2O_5$; (c) Relative enthalpies of $MgSiO_3$ PPv, aggregations of $I$-42$d$-type $Mg_2SiO_4$ and $Fe_2P$-type $SiO_2$, of $P2_1/c$-type $MgSi_2O_5$ and CsCl-type MgO, and of CsCl-type MgO and $Fe_2P$-type $SiO_2$ with respect to an aggregation of $I$-42$d$-type $Mg_2SiO_4$ and $P2_1/c$-type $MgSi_2O_5$.

Here we have searched for phases in an enlarged composition space and considered also the gradual ex-solution of $SiO_2$ from $MgSiO_3$:

**3)** **$MgSiO_3$** (PPv) → **$Mg_2SiO_4$** (new) + **$SiO_2$** ($Fe_2P$ -type)

The adaptive-GA was used to investigate possible dissociation pathways involving stepwise ex-solutions of $SiO_2$ and/or of MgO up to ~ 4TPa. This required predictions of crystal structures for various phases with these compositions and unit cells containing up to 56 atoms (8 formula unit of $Mg_2SiO_4$).

**Table 4.** Structural parameters of *I*-42*d* $Mg_2SiO_4$ and *P*2₁/*c* $MgSi_2O_5$ at 1 TPa

| | | |
|---|---|---|
| $Mg_2SiO_4$ | | |
| Space group | | *I*-42*d* |
| (*a*, *c*) (Å) | | (4.5745, 4.7006) |
| Mg | 8d | (0.1118, 0.2500, 0.1250) |
| Si | 4b | (0.0000, 0.0000, 0.5000) |
| O | 16e | (0.4310, 0.3139, 0.2968) |
| | | |
| $MgSi_2O_5$ | | |
| Space group | | $P2_1/c$ |
| (*a*, *b*, *c*) (Å) | | (4.4358, 4.0308, 6.1897) |
| *β* (°) | | 89.5808 |
| Mg | 4e | (0.0056, 0.5641, 0.1688) |
| Si(1) | 4e | (0.5032, 0.4519, 0.8240) |
| Si(2) | 4e | (0.7569, 0.9309, 0.9917) |
| O(1) | 4e | (0.7385, 0.6749, 0.7195) |
| O(2) | 4e | (0.2678, 0.8453, 0.1942) |
| O(3) | 4e | (0.7263, 0.5331, 0.0000) |
| O(4) | 4e | (0.9575, 0.1890, 0.0776) |
| O(5) | 4e | (0.4606, 0.8275, 0.9139) |

Results confirm the stability of all structures previously identified in the dissociation pathway of $MgSiO_3$ [13, 33, 34]. The intermediate $SiO_2$- and MgO-rich compounds we now include are $MgSi_2O_5$, and $Mg_2SiO_4$. A previous structural search on $Mg_2SiO_5$ was done following soft phonon modes [34]. Here the search for structures is more thorough using the adaptive-GA. MgO undergoes a pressure induced transition from NaCl-type to a CsCl-type phase at ~0.53 TPa, consistent with other first-principles calculations [35-38]. $SiO_2$ undergoes a phase transition from pyrite- to a $Fe_2P$-type phase at ~0.69 TPa [13, 39]. The previously predicted structure of $MgSi_2O_5$ is $P2_1/c$-type [34] and has 32 atoms per cell. The AGA search carried out using up to 48 atoms per cell confirms this $P2_1/c$-type phase to be the lowest-enthalpy phase at the high pressures considered here. Here we predicted a new high-pressure phase of $Mg_2SiO_4$. This structure is body-centered-tetragonal with space group *I*-42*d*. Hereafter, we refer to this phase as *I*-42*d*-type $Mg_2SiO_4$. As far as we know, this structure has not been identified in any substance. However, its cation configuration is identical to that of $Zn_2SiO_4$-II whose space group is *I*-42*d* also [40]. The difference between *I*-42*d*-type $Mg_2SiO_4$ and $Zn_2SiO_4$-II structures is in the oxygen arrangement. *I*-42*d*-type $Mg_2SiO_4$ is much more closely-packed than $Zn_2SiO_4$-II. Both Mg and Si atoms in *I*-42*d*-type $Mg_2SiO_4$ are eight-fold coordinated, while Zn and Si atoms in $Zn_2SiO_4$-II are four-fold coordinated (*i.e.* in tetrahedra). The crystal structures of *I*-42*d*-type $Mg_2SiO_4$ and $P2_1/c$-type $MgSi_2O_5$ together with the corresponding typical local structures are shown in figure 4, and the structural parameters are given in table 4.

The enlarged *composition space* of possible structures, now reveals a more complex three-step dissociation process: $MgSiO_3$ PPv first decomposes into two phases at 0.77 TPa, one rich (poor) and one poor (rich) in $SiO_2$ (MgO):

1) 3 $\mathbf{MgSiO_3}$ (PPv) → $\mathbf{Mg_2SiO_4}$ (*I*-42*d*-type) + $\mathbf{MgSi_2O_5}$ ($P2_1/c$-type)

$MgSi_2O_5$ then breaks down into $Mg_2SiO_4$ and $SiO_2$ at 1.25 TPa:

2) $\mathbf{Mg_2SiO_4}$ (*I*-42*d*-type) + $\mathbf{MgSi_2O_5}$ ($P2_1/c$-type) → 3/2 $\mathbf{Mg_2SiO_4}$ (*I*-42*d*-type) + 3/2 $\mathbf{SiO_2}$ ($Fe_2P$-type)

The complete dissociation into oxides then takes place at 3.09 TPa:

3) 3/2 $\mathbf{Mg_2SiO_4}$ (*I*-42*d*-type) + 3/2 $\mathbf{SiO_2}$ ($Fe_2P$-type) → 3 **MgO** (CsCl-type) + 3 $\mathbf{SiO_2}$ ($Fe_2P$-type)

This sequence of transitions is shown in figure 4(c). The dissociation starts at 0.77 TPa and is completed at ~3TPa. This pressure range is expected to change somewhat at high temperatures though. This prediction contrasts with the direct dissociate previously predicted at 1.21 TPa [33] and the two-step ex-solution of MgO predicted between 0.90 and 2.1 TPa [34]. In overall, these results suggest that more complex sequences of phase transitions could be found by considering intermediate phases with different compositions. The large ionic (LDA) gaps of at least 5 eV found in all phases suggest that phases with different molar fractions of MgO and $SiO_2$ with even more complex structures are the most natural candidates for future investigations.

In summary, we showed that the adaptive-GA is an efficient scheme for complex crystal structure prediction. The use of adjustable classical potentials in the internal loop of the adaptive-GA improves the speed and efficiency of the search by several orders of magnitude without compromising the DFT accuracy of the final result. Its performance was demonstrated in studies of complex binary and ternary intermetallic alloys. The pressure induced dissociation of $MgSiO_3$, the major end-member phase of the Earth mantle, into elementary oxides at TPa pressures was also investigated. A three-step dissociation process happening between 0.44 and 3.09 TPa involving complex crystal structures and compositions was identified. Structure prediction in the Mg-Si-O systems is an outstanding example of materials discovery at extreme conditions of planetary interiors where a wealth of novel phases with exotic properties [33] is expected to exist. The adaptive-GA combined with peta-scale computing power is an invaluable method to search for these phases.

Work at Ames Laboratory was supported by the US Department of Energy, Basic Energy Sciences, Division of Materials Science and Engineering, under Contract No. DE-AC02-07CH11358, including a grant of computer time at the National Energy Research Supercomputing Centre (NERSC) in Berkeley, CA. Work at University of Minnesota was supported by NSF/EAR 1047629. Computations were performed at the Minnesota Supercomputing Institute and at the Laboratory for Scientific Computing and Engineering. SQW also acknowledges financial support from the National Natural Science Foundation of China (No. 11004165).